\documentclass[twocolumn,aps,showpacs,floatfix,superscriptaddress]{revtex4}
\usepackage{amsmath,amssymb,eucal,graphicx}
\begin{document}
\title{Dynamics of Social Balance on Networks}
\author{T.~Antal}
\altaffiliation{On leave from Institute for Theoretical Physics -- HAS,
  E\"otv\"os University, Budapest, Hungary} 
\author{P. L. Krapivsky}
\affiliation{Department of Physics, Boston University, Boston, Massachusetts
02215, USA}
\author{S.~Redner}
\email{redner@bu.edu}
\altaffiliation{Permanent address:
Department of Physics, Boston University, Boston, Massachusetts
02215} \affiliation{Theoretical Division and Center for
Nonlinear Studies, Los Alamos National Laboratory, Los Alamos, New
Mexico 87545, USA}

\begin{abstract}
  
  We study the evolution of social networks that contain both friendly and
  unfriendly pairwise links between individual nodes.  The network is endowed
  with dynamics in which the sense of a link in an imbalanced triad---a
  triangular loop with 1 or 3 unfriendly links---is reversed to make the
  triad balanced.  With this dynamics, an infinite network undergoes a
  dynamic phase transition from a steady state to ``paradise''---all links
  are friendly---as the propensity $p$ for friendly links in an update event
  passes through 1/2.  A finite network always falls into a socially-balanced
  absorbing state where no imbalanced triads remain.  If the additional
  constraint that the number of imbalanced triads in the network does not
  increase in an update is imposed, then the network quickly reaches a
  balanced final state.

\end{abstract}


\maketitle

\section{Introduction}

In this work, we investigate the role of friends and enemies on the evolution
of social networks.  We represent individuals as nodes in a graph and a
relationship between individuals as a link that joins the corresponding
nodes.  To quantify a relationship, we assign the binary variable $s_{ij}=\pm
1$ to link $ij$, with $s_{ij}=1$ if nodes $i$ and $j$ are friends, and
$s_{ij}=-1$ if $i$ and $j$ are enemies (Fig.~\ref{process}).  A basic
characterization of relationships between mutual acquaintances is the notion
of {\em social balance} \cite{FH,social}.  The triad $ijk$ is defined as
balanced if the sign of the product of the links in the triad
$s_{ij}s_{jk}s_{ki}$ equals 1, while the triad is imbalanced otherwise.  We
define a triad to be of type $\triangle_k$ if it contains $k$ negative links.
Thus $\triangle_0$ and $\triangle_2$ are balanced, while $\triangle_1$ and
$\triangle_3$ are imbalanced.  A balanced triad therefore fulfills the adage
\begin{itemize}
\itemsep -4pt
\item a friend of my friend is my friend;
\item an enemy of my friend is my enemy;
\item a friend of my enemy is my enemy;
\item an enemy of my enemy is my friend,
\end{itemize}
while an imbalanced triad is analogous to a frustrated plaquette in a random
magnet \cite{SG}.

A network is balanced if each constituent triad is balanced \cite{FH,social}.
An ostensibly more general definition of a balanced network is that {\em
  every} cycle in the network is balanced.  Cartwright and Harary showed
\cite{H} that the cycle-based and triad-based definitions of balance are
equivalent on complete graphs.  Their result implies that if an imbalanced
cycle of any length exists in a complete graph, an imbalanced triad also
exists.

Balance theory has been initiated by Heider \cite{FH} and other social
psychologists \cite{Lewin,Newcomb}, and the subject remains an active
research area \cite{Leik,social,B,Davis,Hum,Dor,HD}.  Much of this work was
devoted to classifying balanced stable states of networks when relationships
are static.  A fundamental result from these studies is that balanced
societies are remarkably simple: either all individuals are mutual friends
(we call such a state ``paradise''), or the network segregates into two
antagonistic cliques where individuals within the same clique are mutual
friends and individuals from distinct cliques are enemies (we call such a
state ``bipolar'') \cite{H}.  Balance theory also has natural applications to
international relations \cite{Moore}.  As a particularly compelling example,
the Triple Alliance (1882) pitted Germany, Austria-Hungary, and Italy against
the Triple Entente (1907) countries of Britain, France, and Russia \cite{L}.
This bipolar state of competing alliances clearly contributed to the onset of
World War I.

A large network is almost surely imbalanced if the relationships are randomly
chosen to be friendly or unfriendly.  Clearly such a network is socially
unstable and the web of relations must evolve to a more stable state if the
individual nodes behave rationally.  In this work, we go beyond a static
description social relations and investigate how an initially imbalanced
society achieves balance by endowing a network with a prototypical social
dynamics that reflects the natural human tendency to reduce imbalanced
triads.  A related line of investigation, based on evolving social networks
with continuous interaction strengths, has also recently appeared \cite{K}.

\section{Models}

\begin{figure}[ht]
  \includegraphics[width=0.9\linewidth]{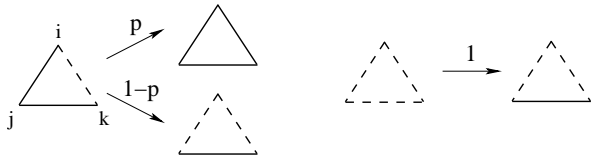}
  \caption{Imbalanced triads $\triangle_1$ (left) and $\triangle_3$ (right) 
    and the possible outcomes after an update step by local triad dynamics.
    Full and dashed lines represent friendly ({\it e.g.}, $s_{ij}=1$) and
    unfriendly ({\it e.g.}, $s_{ik}=-1$) relations respectively.  }
\label{process}
\end{figure}

We first consider what we term {\em local triad dynamics} (LTD).  In an
update step of LTD, we first choose a triad at random.  If this triad is
balanced ($\triangle_0$ or $\triangle_2$), no evolution occurs.  If the triad
is imbalanced ($\triangle_1$ or $\triangle_3$), we change $s\to -s$ on one of
the links as follows: $\triangle_1\to \triangle_0$ occurs with probability
$p$, and $\triangle_1\to \triangle_2$ occurs with probability $1-p$, while
$\triangle_3\to \triangle_2$ occurs with probability 1 (Fig.~\ref{process}).
One unit of time is defined as $L$ update events, where $L$ is the total
number of links.  Notice that for the special case of $p=1/3$, each link of
an imbalanced triad is flipped equiprobably.  When $p>1/3$, the density of
friendly links consequently tends to increase and the society is predisposed
to tranquility, while for $p<1/3$ the societal predisposition is hostility.

After an update step in LTD, the imbalanced target triad becomes balanced,
but other balanced triads that share a link with this target may become
imbalanced.  These triads can subsequently evolve and return to balance,
leading to new imbalanced triads.  Such an interaction cascade is familiar in
social settings.  For example, if a married couple breaks up, the
acquaintances of the couple may then be obliged to redefine their relations
with each partner in the couple so as to maintain balanced triads.  These
redefinitions, may lead to additional relationship shifts, {\it etc}.

For $p<1/2$, we shall show that LTD quickly drives an infinite network to a
quasi-stationary dynamic state where global characteristics, such as the
densities of friendly relations or imbalanced triads, fluctuate around
stationary values.  As $p$ passes through a critical value of $1/2$, the
network undergoes a phase transition to a paradise state where no unfriendly
relations remain.  On the other hand, a finite network always reaches a
balanced state.  For $p<1/2$, this balanced state is bipolar and the time to
reach this state scales faster than exponentially with network size.  For
$p\geq 1/2$, the final state is paradise.  The time to reach this state
scales algebraically with $N$ when $p=1/2$, and logarithmically in $N$ for
$p>1/2$.

We also investigate {\em constrained triad dynamics} (CTD).  Here, we select
a random link, and change $s\to -s$ for this link if the total number of
imbalanced triads decreases.  If the total number of imbalanced triads is
conserved in an update, then the update occurs with probability 1/2.  Updates
that would increase the total number of imbalanced triads are not allowed.
We again define the unit of time as $L$ update events, so that on average
each link is changed once in unit of time.  The global constraint accounts
for the socially-plausible feature that an agent considers all of its mutual
acquaintances before deciding to change the character of a relationship.  CTD
also corresponds to an Ising model with a three-spin interaction between the
links of a triad, $\mathcal{H}= -\sum_{ijk}s_{ij}s_{jk}s_{ki}$, where the sum
is over all triads $ijk$, with zero-temperature Glauber dynamics
\cite{glauber}.  As we shall see, a crucial outcome of CTD is that a network
is quickly driven to a balanced state in a time that scales as $\ln N$.

In the following two sections we analyze the dynamics of networks that evolve
by LTD or CTD.  For simplicity, we consider networks with a complete graph
topology---everyone knows everyone else.  This limit is appropriate for small
networks, such as the diplomatic relations of countries.  We then summarize
and discuss some practical implications of our results in Sec.~IV.

\section{Local Triad Dynamics}

\subsection{Evolution Equations}

We begin with essential preliminaries for writing the governing equations for
the various triad densities.  Let $N$, $L=\binom{N}{2}$, and $N_{\triangle} =
\binom{N}{3}$ be the numbers of nodes, links, and triads in the network.
Define $N_k$ as the number of triads that contain $k$ negative links, with
$n_k=N_k/N_\triangle$ the respective triad densities, and $L^+$ ($L^-$) the
number of positive (negative) links.  The number of triads and links are
related by
\begin{equation}
L^+ \!=\! \frac{3N_0+2N_1+N_2}{N-2}\,,\quad
L^-\!=\!\frac{N_1+2N_2+3N_3}{N-2}\,\,.
\label{L+-}
\end{equation}
The numerator counts the number of positive links in all triads while the
denominator appears because each link is counted $N-2$ times.  The density of
positive links is therefore $\rho=L^+/L=(3n_0+2n_1+n_2)/3$, while the density
of negative links is $1-\rho=L^-/L$.

A fundamental network characteristic is $N_k^+$, which is defined as follows:
for each positive link, count the number of triads of type $\triangle_k$ that
are attached to this link.  Then $N_k^+$ is the average number of such triads
over all positive links.  This number is
\begin{equation}
\label{Ni+}
N_k^+ = \frac{(3-k)N_k}{L^+}.
\end{equation}
The factor $(3-k)N_k$ accounts for the fact that each of the $N_k$ triads of
type $\triangle_k$ is attached to $3-k$ positive links; dividing by $L^+$
then gives the average number of such triads.  Analogously, we introduce
$N_k^-=kN_k/L^-$.  Since the total number of triads attached to any given
link equals $N-2$, the corresponding triad densities are
\begin{subequations}
\begin{align}
&n_k^+=\frac{N_k^+}{N-2} = \frac{(3-k)n_k}{3n_0+2n_1+n_2}
\label{ni+}\\
&n_k^-=\frac{N_k^-}{N-2} = \frac{kn_k}{n_1+2n_2+3n_3} ~.
\label{ni-}
\end{align}
\end{subequations}

We now write the rate equations that account for changes in the various triad
densities in a single update event.  We choose a triad at random; if it is
imbalanced ($\triangle_1$ or $\triangle_3$) we change one of its links as
shown in Fig.~\ref{process}.  Let $\pi^+$ be the probability that a link
changes from $+$ to $-$ in an update event, and vice versa for $\pi^-$.
Since a link changes from $1\to -1$ with probability $1-p$ when
$\triangle_1\to \triangle_2$, while a link changes from $-1\to 1$ with
probability $p$ if $\triangle_1\to \triangle_0$ and with probability 1 if
$\triangle_3\to \triangle_2$, we have (see Fig.~\ref{process})
\begin{equation}
\label{pi}
\pi^+ = (1-p)\,n_1 \qquad \pi^- = p\,n_1+n_3.
\end{equation}

Since each update changes $N-2$ triads, and we also defined one time step as
$L$ update events, the rate equations for the triad densities have the
size-independent form
\begin{equation}
\label{ni-rate}
\begin{split}
&\frac{dn_0}{dt}  =  \pi^- n_1^- - \pi^+ n_0^+\,,\\ 
&\frac{dn_1}{dt}  =  \pi^+ n_0^+ + \pi^- n_2^- - \pi^- n_1^- - \pi^+n_1^+\,,\\
&\frac{dn_2}{dt}  =  \pi^+ n_1^+ + \pi^- n_3^- - \pi^- n_2^- - \pi^+n_2^+\,,\\
&\frac{dn_3}{dt}  =  \pi^+ n_2^+ - \pi^- n_3^-\,.
\end{split}
\end{equation}

\subsection{Stationary States}
\label{SS} 

We first study stationary states.  Setting the left-hand sides of
Eqs.~(\ref{ni-rate}) to zero and also imposing $\pi^+=\pi^-$ to ensure a
fixed friendship density, we obtain
\begin{equation*}
n_0^+ = n_1^-~,~~ n_1^+ = n_2^-~,~~ n_2^+=n_3^- \,.
\end{equation*}
By forming products such as $n_0^+n_2^-=n_1^+n_1^-$, these relations are
equivalent to
\begin{equation}
\label{stati} 
3n_0 n_2 = n_1^2~,\qquad 3n_1 n_3 = n_2^2\,. 
\end{equation}

\begin{figure}[ht]
  \includegraphics[width=1.0\linewidth]{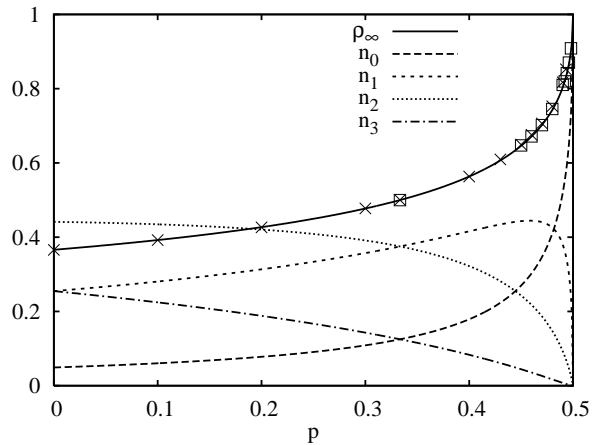}
  \caption{Exact stationary densities $n_k(p)$ and the density of friendly
    relations $\rho_\infty$ as a function of $p$.  Simulation results for
    $\rho_\infty$ for $N=64$ (crosses) and 256 (boxes) are also shown.}
\label{n-vs-p}
\end{figure}

Substituting $\pi^+$ and $\pi^-$ from Eq.~(\ref{pi}) into the stationarity
condition, $\pi^+=\pi^-$, gives $n_3=(1-2p)n_1$.  Using this result, as well
as the normalization $\sum n_k=1$, in Eqs.~(\ref{stati}), we find, after some
straightforward algebra
\begin{equation}
\label{stat-nj}
n_j=\binom{3}{j} \rho_\infty^{3-j}(1-\rho_\infty)^{j},
\end{equation}
where 
\begin{equation}
\label{stat-friends}
\rho_\infty=
\begin{cases}
1/[\sqrt{3(1-2p)}+1]   &\qquad    p\leq 1/2;\cr       
1                      &\qquad    p\geq 1/2;\cr
\end{cases}
\end{equation}
is the stationary density of friendly links. Equation (\ref{stat-nj}) shows
that relationships are {\em uncorrelated\/} in the stationary state.  As
shown in Fig.~\ref{n-vs-p}, the density of friendly links $\rho_\infty$
monotonically increases with $p$ for $0\leq p\leq 1/2$, while for $p\geq
1/2$, paradise is reached where all people are friends.  Near the phase
transition, the density of unfriendly links $1-\rho_\infty$ vanishes as
$\sqrt{3\epsilon}+O(\epsilon)$, as $\epsilon\equiv 1-2p\to 0$.

\subsection{Temporal Evolution} 
\label{evol} 

A remarkable feature of equations (\ref{ni-rate}) is that if the initial
triad densities are given by Eq.~(\ref{stat-nj})---uncorrelated
densities---the densities will remain uncorrelated forever.  For such initial
conditions it therefore suffices to study the time evolution of the density
of friendly links $\rho(t)$.  This time evolution can be extracted from
Eqs.~(\ref{ni-rate}), or it can be established directly by noting that
$\rho(t)$ increases if $\triangle_3\to \triangle_2$ or $\triangle_1\to
\triangle_0$, and decreases if $\triangle_1\to \triangle_2$.  Taking into
account that the respective probabilities for these processes are $1, p$, and
$1-p$, we find
\begin{equation}
\label{rho}
\frac{d\rho}{dt}=3(2p-1)\rho^2(1-\rho)+(1-\rho)^3.
\end{equation}
Thus the time dependence of the density of friendly links is given by the
implicit relation
\begin{equation}
\label{rho-sol}
\int_{\rho_0}^\rho \frac{dx}{3(2p-1)x^2(1-x)+(1-x)^3}=t.
\end{equation}

When $p<1/2$, the stationary density of Eq.~(\ref{stat-friends}) is
approached exponentially in time:
\begin{equation*}
\rho(t)-\rho_\infty \sim e^{-Ct}~,~~ C=\frac{6\epsilon}{1+\sqrt{3\epsilon}} ~,
\end{equation*}
where again $\epsilon=1-2p$.  At the threshold value $p=1/2$, the friendship
density is given by
\begin{equation}
\label{rho-sol-crit}
\rho=1-\frac{1-\rho_0}{\sqrt{1+2(1-\rho_0)^2t}} ~.
\end{equation}
Here the approach to paradise is algebraic in time, {\it viz.}, $1-\rho\to
1/\sqrt{2t}$ as $t\to\infty$.  As a consequence,
\begin{equation}
\label{n123}
(n_0,n_1,n_2,n_3)\to \left(1-\frac{3}{\sqrt{2t}}\,,\frac{3}{\sqrt{2t}}\,,
\frac{3}{2t}\,,\frac{1}{(2t)^{3/2}}\right).
\end{equation}
Finally when $p>1/2$,  
\begin{equation}
\label{appr-to-paradise}
1-\rho\sim \exp\left[-3(2p-1)t\right] ~,
\end{equation}
so that paradise is reached exponentially quickly. 

\subsection{Fate of a Finite Society} 
\label{fate} 

Although an infinite network reaches a dynamic steady state for $p<1/2$, a
finite network ultimately falls into an absorbing state for all $p$.  Such
absorbing states are necessarily balanced, because any network that contains
an imbalanced triad continues to change.  To see why such an absorbing state
must eventually be reached, consider the evolution in which {\em at each
  step} an unfriendly link changes to friendly in an imbalanced triad.  Since
the number of unfriendly links always decreases, a balanced state is reached
in a finite number steps.  Finally, because this particular route to a
balanced state has a nonzero probability to occur, any initial network
ultimately reaches an absorbing balanced state.

\begin{figure}[htb]
\hskip -0.28in     \includegraphics[width=1.08\linewidth]{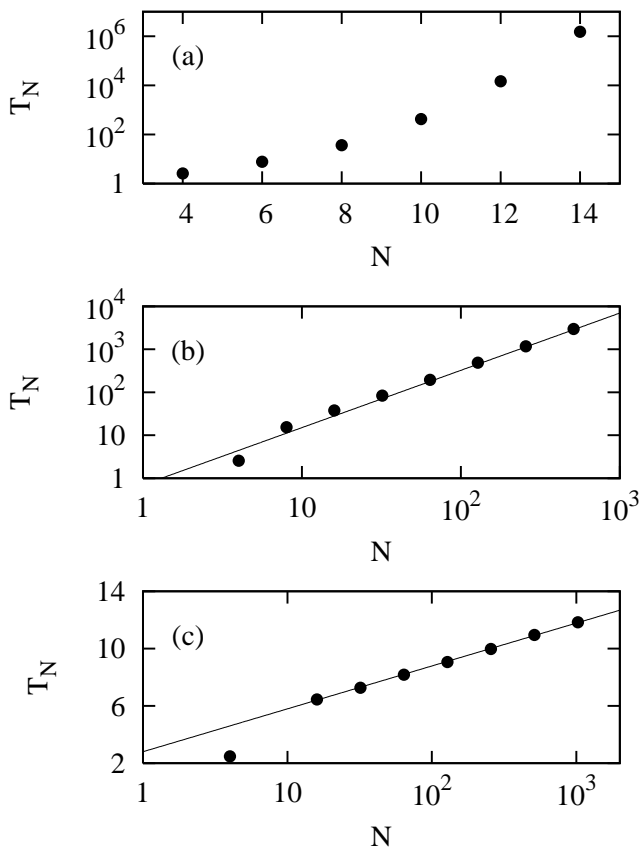} 
     \caption{Average absorption time as a function of $N$ for an initially
       antagonistic society ($\rho_0=0$) for: (a) $p=1/3$; (b) $p=1/2$; (c)
       p=3/4.  The line in (b) has slope $4/3$. }
\label{avtime}
\end{figure}

Our simulations show that a finite network evolves to a bipolar state for
$p<1/2$, independent of the initial state.  The size difference of the two
final cliques is virtually independent of the initial configuration because
the network spends an enormous time in a quasi-stationary state before
reaching the absorbing state.  For $p<1/2$, we estimate the time to reach a
bipolar state by the following crude argument \cite{fpp}.  Consider a nearly
balanced network.  When a link is flipped on an imbalanced triad, then of the
order of $N$ new imbalanced triads will be created in the adjacent triads
that contain the flipped link.  Thus starting near a balanced state, local
triad dynamics is equivalent to a biased random walk in the state space of
all network configurations, in which the bias is directed away from the
balanced state, with the bias velocity $v$ proportional to $N$.  Conversely,
when the network is far from balance, local triad dynamics is diffusive in
character because the number of imbalanced triads will change by of the order
of $\pm N$ equiprobably in a single update.  The corresponding diffusion
coefficient $D$ is then proportional to $N^2$.  Since the total number of
triads in a network of $N$ nodes is $N_\triangle\sim N^3$, we therefore
expect that the time $T_N$ to reach balance will scale as
\begin{equation}
\label{bounds}
T_N\sim e^{v N_\triangle/D}\sim e^{N^2}.
\end{equation}

When $p\geq 1/2$, paradise is reached with a probability that quickly
approaches one as $N\to\infty$.  At the threshold $p=1/2$, a naive estimate
for the time $T_N$ to reach paradise is given by the time at which the
density of unfriendly links $a(t)\equiv 1-\rho(t)$ is of the order of
$N^{-2}$, corresponding to one unfriendly link in the network.  From
Eq.~(\ref{rho-sol-crit}), the criterion $a(T_N)\sim N^{-2}$ gives $T_N\sim
N^4$.  While simulations show that $T_N$ does scales algebraically with $N$,
the exponent value is much smaller (Fig.~\ref{avtime}(b)).  The source of
this smaller exponent is the existence of anomalously large fluctuations in
the number of unfriendly links.

To determine these fluctuations in the thermodynamic limit, we write the
number of unfriendly links $A(t)\equiv L^-(t)$ in the canonical form
\cite{vK}
\begin{equation}
\label{expansion}
A(t)=La(t)+\sqrt{L}\,\eta(t),
\end{equation}
where $a(t)$ is deterministic and $\eta(t)$ is a stochastic variable.  Both
$a$ and $\eta$ are size independent in the thermodynamic limit ($L\gg 1$),
and the form of Eq.~(\ref{expansion}) assures that the average $\langle
A\rangle$ and the variance $\langle A^2\rangle - \langle A\rangle^2$ grow
linearly with the total number of links $L$.  In Appendix~\ref{fluctsig}, we
show that $\sigma\equiv \langle\eta^2\rangle$ grows as
\begin{equation}
\label{sigma}
\sigma\sim \sqrt{t} \qquad {\rm as}\quad t\to\infty ~.
\end{equation}

Thus the time to reach paradise $T_N$ is determined by the criterion that
fluctuations in $A$ become of the same order as the average, {\it viz.},
\begin{equation}
\label{criterion}
 \sqrt{L\sigma(T_N)}\sim L a(T_N) ~.
\end{equation}
Using $a(t)\sim 1/\sqrt{t}$ from Eq.~(\ref{rho-sol-crit}) together with
Eq.~(\ref{sigma}) and $L\sim N^2$, we rewrite Eq.~(\ref{criterion})
as $N^2\, T_N^{-1/2}\sim N\, T_N^{1/4}$.  This leads to the estimate
\begin{equation}
\label{correct}
T_N\sim N^{4/3}\,. 
\end{equation}

Above the threshold $p>1/2$, paradise is approached exponentially
quickly [see Eq.~(\ref{appr-to-paradise})] and the time to paradise scales
logarithmically with network size:
\begin{equation}
\label{TN-above}
T_N\sim (2p-1)^{-1}\,\ln N
\end{equation}
Interestingly, the estimates (\ref{correct}) and (\ref{TN-above}) coincide
when $2p-1\sim N^{-4/3}\,\ln N$.  That is, there is a finite-width critical
region near the phase transition due to finite-size effects.  In
Appendix~\ref{fluctsig}, we estimate this width by analyzing the fluctuations
below the threshold $p<1/2$ and obtain essentially this same result.

Summarizing, the asymptotics for the absorption time are:
\begin{equation}
\label{BM-T}
T_N\propto 
\begin{cases}
\exp\left(N^2\right)        &p<1/2\\
N^{4/3}                     &p=1/2\\
(2p-1)^{-1}\,\ln N          &p>1/2 
\end{cases}
\end{equation}
in agreement with the simulation results in Fig.~\ref{avtime}.

\section{Constrained Triad Dynamics}

\subsection{Jamming and Absorption Time} 
\label{sim}

\begin{figure}[!h]
     \includegraphics[width=1.0\linewidth]{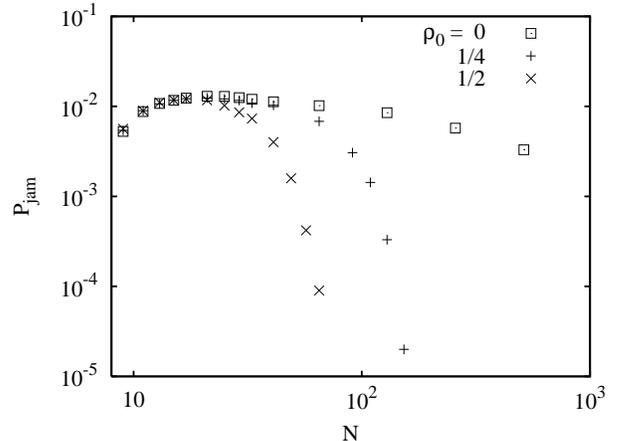} 
     \caption{Probability of reaching a jammed state $P_\mathrm{jam}$ as a 
     function of $N$ for several values of $\rho_0$. }
\label{Bpsi}
\end{figure}

In constrained triad dynamics (CTD), the number of imbalanced triads cannot
increase in an update event, and the final state can either be balanced or
jammed.  A jammed state is one in which imbalanced triads exist and for which
the flip of any link increases the number of imbalanced triads.  Since this
type of update is forbidden in CTD, there is no escape from a jammed state.
Moreover, jammed states turn out to be much more numerous than balanced
states (see Sec.~\ref{enumer}).  In spite of this fact, we find that the
probability for the network to reach a jammed state, $P_\mathrm{jam}(N)$,
quickly goes to zero as $N$ increases, except for the case of an initially
antagonistic society ($\rho_0=0$), where $P_\mathrm{jam}(N)$ decays slowly
with $N$ (Fig.~\ref{Bpsi}).  Thus the final network state is almost always
balanced for large $N$, and consists either of one clique (paradise) or two
antagonistic cliques.  It is worth mentioning that we never observed
``blinkers'' \cite{SKR} in simulations, although we cannot prove that such
states do not exist.  These are trajectories in the state space that evolve
forever and correspond to a network in which all possible updates involve no
change in the number of imbalanced triads.

Another fundamental feature of CTD is that the time $T_N$ for a network of
$N$ nodes to reach its final state generically scales as $\ln N$.  While
$T_N$ now depends on the initial condition, in contrast to LTD, this
dependence occurs either in the amplitude or in lower-order corrections of
the absorption time.  Thus the logarithmic growth of $T_N$ with $N$ is a
robust feature of CTD.

\subsection{Final Clique Sizes}

\begin{figure}[htb]
     \includegraphics[width=1.0\linewidth]{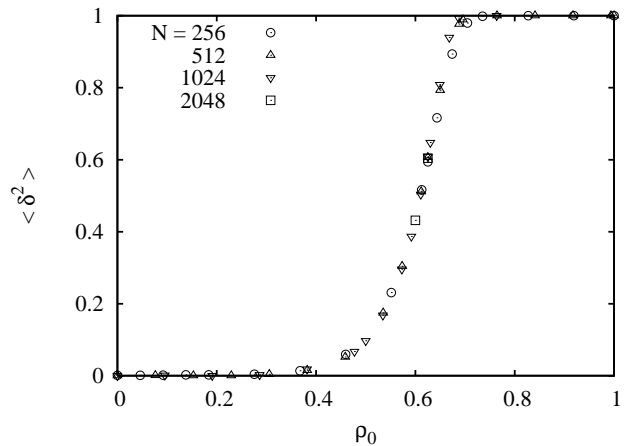} 
     \caption{Asymmetry of the final state as a function of the
       initial friendship density $\rho_0$ for several network sizes.}
\label{BavDi}
\end{figure}

\begin{figure}[htb]
     \includegraphics[width=1.0\linewidth]{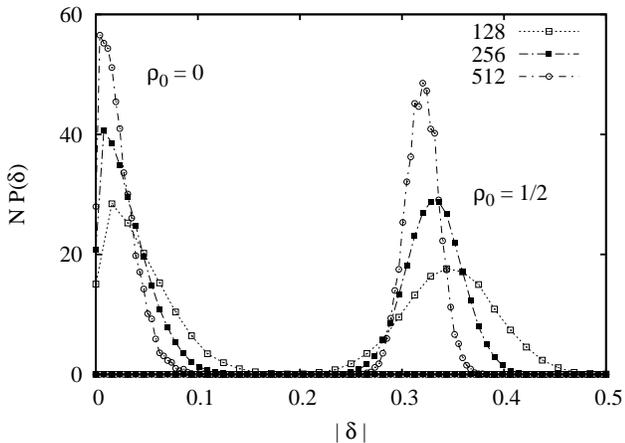} 
     \caption{Scaled probability distribution of the relative difference between the
       final clique sizes for $\rho_0=0$ and $1/2$. }
\label{BPd}
\end{figure}

An unexpected feature of CTD is the phase transition for the difference in
sizes $C_1$ and $C_2$ of the two final cliques as a function of $\rho_0$
(Fig.~\ref{BavDi}).  We quantify this asymmetry by the scaled size difference
$\delta=(C_1-C_2)/N$.  For $\rho_0\alt 0.4$ the cliques sizes in the final
bipolar state are nearly the same size and $\langle\delta^2\rangle\approx 0$.
As $\rho_0$ increases toward $\rho_0^* \approx 0.65$, the size difference of
the two cliques continuously increases.  A sudden change occurs at
$\rho_0^*$, beyond which the final network state is paradise.  The
probability distribution for $\delta$ is sharply peaked about its average
value as $N\to\infty$ (Fig.~\ref{BPd}).  Since $\langle \delta^2\rangle$ and
the density of friendly links $\rho_\infty$ are related by $\langle \delta^2
\rangle = 2\rho_\infty-1$ in a large balanced society, uncorrelated initial
relations generically lead to $\rho_\infty>\rho_0$.  Thus CTD tends to drive
a network into a friendlier final state.

While we do not have a detailed understanding of this phase transition, we
give a qualitative argument that suggests that a large network undergoes a
sudden change from $\rho_\infty=0$ (two equal size cliques) when $\rho_0<1/2$
to $\rho_\infty=1$ (paradise) when $\rho_0>1/2$.  The fact that the
transition appears to be located near $\rho^*_0\approx 0.65$
(Fig.~\ref{BavDi}) rather than at $\rho_0=1/2$ indicates that our approach is
not a complete description for the transition.

We first assume, as observed in simulations of large networks, that jammed
states do not arise.  We also assume that a network remains uncorrelated
during its early stages of evolution.  Consequently the densities ${\bf
  n^+}\equiv (n_0^+,n_1^+,n_2^+,n_3^+)$ of triads that are attached to a
positive link are
\begin{equation}
\label{positive} 
{\bf n^+}=(\rho^2,2\rho(1-\rho),(1-\rho)^2,0).
\end{equation}
For a link to change from $+$ to $-$, it is necessary that $n_1^++n_3^+ >
n_0^++n_2^+$.  From Eq.~(\ref{positive}), this condition is equivalent to
$4\rho(1-\rho)>1$, which never holds.  Similarly, the densities ${\bf
  n^-}\equiv(n_0^-,n_1^-,n_2^-,n_3^-)$ of triads attached to a negative link
are
\begin{equation}
\label{negative} 
{\bf n^-}=(0,\rho^2,2\rho(1-\rho),(1-\rho)^2).
\end{equation}
The requirement $n_1^-+n_3^->n_0^-+n_2^-$ now reduces to $1>4\rho(1-\rho)$,
which is valid when $\rho\ne 1/2$.

Thus for a large uncorrelated network, only negative links flip in CTD.
Since, the density of negative links is $1-\rho$, the governing rate equation
is
\begin{equation}
\label{rho-CTD} 
\frac{d \rho}{d t}=1-\rho,
\end{equation}
from which
\begin{equation}
\label{rho-CTD-sol} 
\rho=1-(1-\rho_0)e^{-t}.
\end{equation}
{}From the criterion $1-\rho(T_N)\sim N^{-2}$, corresponding to one unfriendly
link remaining in the network, the time to reach paradise is given by
$T_N\sim\ln N$, in agreement with simulations.  According to
Eq.~(\ref{rho-CTD-sol}) a network should evolve to paradise for any initial
condition.

However, our simulations indicate that this homogeneous solution is unstable
for $\rho_0<1/2$.  In this case, the density of friendly links $\rho$
initially still increases according to (\ref{rho-CTD-sol}) until $\rho\approx
1/2$.  At this point, correlations in the relationship structure begin to
arise and these ultimately lead to a bipolar society with $\rho_\infty\approx
1/2$.  We now give a qualitative argument to support these observations.

When $\rho(t)\approx 1/2$, there are many partitions of the network into two
subnetworks $\mathcal{S}_1$ and $\mathcal{S}_2$ of nearly equal sizes
$C_1=|\mathcal{S}_1|$ and $C_2=|\mathcal{S}_2|$, for which the density of
friendly links within each subnetwork, $\rho_1$ and $\rho_2$, slightly exceed
$1/2$, while the density $\beta$ of friendly links between subnetworks is
slightly less than $1/2$.  Our basic point is that this small fluctuation is
amplified by CTD so that the final state is two nearly equal-size cliques.

To appreciate how such an evolution can occur, we assume that relationships
within each subnetwork and between subnetworks are homogeneous.  Consider a
negative link in $\mathcal{S}_1$.  The densities of triads attached to this
link are given by (\ref{negative}), with $\rho$ replaced by $\beta$ when the
third vertex in the triad belongs to $\mathcal{S}_2$, and by
(\ref{negative}), with $\rho$ replaced by $\rho_1$ when the third vertex
belongs to $\mathcal{S}_1$.  The requirement that a link can change from $-$
to $+$ according to CTD now becomes
\begin{equation}
\label{req1} 
C_1[1-4\rho_1(1-\rho_1)]+C_2[1-4\beta(1-\beta)]>0,
\end{equation}
which is always satisfied.  Additionally, negative links within each
subnetwork can change to positive with rate one, while positive links within
each subnetwork can never change.

Consider now a positive link between the subnetworks.  The triad densities
attached to this link are given by
\begin{equation*} 
{\bf n}^+_j=(\beta\rho_j,\beta(1-\rho_j)+\rho_j(1-\beta),
(1-\beta)(1-\rho_j),0)
\end{equation*}
when the third vertex belongs to $\mathcal{S}_j$.  Since
\begin{equation*}
  \beta(1-\rho_j)+\rho_j(1-\beta)-\beta\rho_j-(1-\beta)(1-\rho_j)=
  (2\rho_j-1)(1-2\beta)\,,
\end{equation*}
the change $+\to -$ is possible if
\begin{equation}
\label{req2} 
[C_1(2\rho_1-1)+C_2(2\rho_2-1)](1-2\beta)>0\,.
\end{equation}
Thus if the situation arises where $\rho_1>1/2$, $\rho_2>1/2$, and
$\beta<1/2$, the network subsequently evolves to increase the density of
intra-subnetwork friendly links and decrease the density of inter-subnetwork
friendly links.  These link densities thus evolve according to the rate
equations
\begin{equation}
\label{rate} 
\frac{d \rho_1}{d t}=1-\rho_1\,,\qquad
\frac{d \rho_2}{d t}=1-\rho_2\,,\qquad
\frac{d \beta}{d t}=-\beta\,,
\end{equation}
and give the instability needed to drive the network to a final bipolar
state.

The last step in our argument is to note that when $C_1\approx C_2\approx
N/2$, the number of ways, $N\choose C_1$, to partition the original network
into the two nascent subnetworks $\mathcal{S}_1$ and $\mathcal{S}_2$, is
maximal.  Consequently, the partition $C_1=C_2$ has the highest likelihood of
providing the initial link density fluctuation, after which the homogeneous
evolution (\ref{rho-CTD}) is replaced by the clique evolution (\ref{rate}) so
that a homogeneous network organizes into two nearly equal-size cliques.
Although our argument fails to account for the quantitative details of the
transition shown in Fig.~\ref{BavDi}, the primary behaviors of $\langle
\delta^2\rangle$ in the two limiting cases of $\rho_0\to 0$ and $\rho_0\to 1$
are described correctly.

\subsection{Structure of Jammed Configurations} 
\label{construct}

\begin{figure}[htb]
\includegraphics[width=1.3\linewidth]{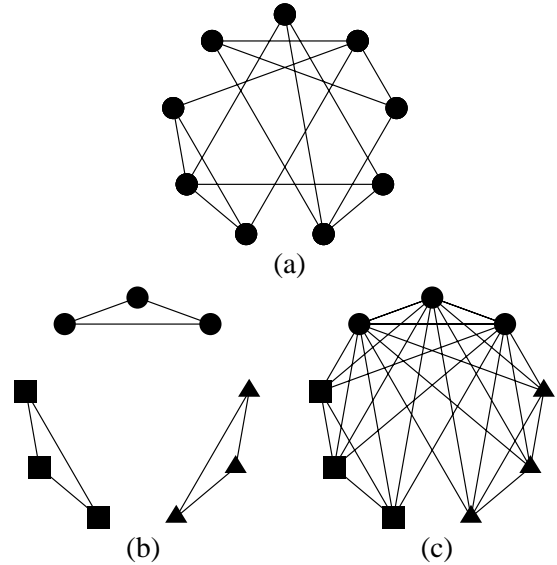}
\caption{Examples of jammed configurations for $N=9$ (only friendly links are
  displayed).  (a) A jammed configuration that appeared in simulations.  (b)
  A jammed state consisting of three mutually antagonistic cliques.  (c) A
  jammed state derived from (b) in which the top clique from (b) is friendly
  toward the remaining two cliques. }
\label{strange9z}
\end{figure}

While jammed configurations can arise in CTD, we will now show that: {\em
  Jammed states are possible if and only if the network size is} $N=9$ {\em
  or} $N\geq 11$.  To prove this statement, we first explicitly construct
jammed configurations for $N=9$ and $N\ge11$.  Fig.~\ref{strange9z} shows
three jammed configurations for $N=9$, the smallest possible $N$ where jammed
configurations can occur.  The example in Fig.~\ref{strange9z}(a) was
observed in simulations, while the jammed configuration in
Fig.~\ref{strange9z}(b) consists of three antagonistic cliques of 3 nodes
each.  We now generalize this latter construction of jammed states to
arbitrary $N\ge11$.

Consider three mutually antagonistic cliques of sizes $(m_1,m_2,m_3)$, with
$m_1+m_2+m_3=N$.  A link within a clique is necessarily stable, as all
attached triads are of type $\triangle_0$ or $\triangle_2$.  Conversely, a
negative link between clique 1 (circles in Fig.~\ref{strange9z}(b)) and clique
2 (squares) is attached to both stable and imbalanced triads.  There are
$m_1-1+m_2-1$ attached stable triads of type $\triangle_2$, where the third
node of the triad is either within clique 1 or clique 2, and $m_3$ attached
imbalanced triads of type $\triangle_3$, where the third node is in clique 3
(triangles).  The requirement for link stability among these cliques is then
\begin{eqnarray}
\label{m123}
&&m_1+m_2>m_3+2\nonumber\\
&&m_2+m_3>m_1+2\\
&&m_3+m_1>m_2+2,\nonumber
\end{eqnarray}
where the last two equations arise by cyclic permutations of the first.

We term a partition $(m_1,m_2,m_3)$ ``jammed'' if it satisfies the
inequalities (\ref{m123}).  By summing pairs of Eqs.~(\ref{m123}), we find
$m_j\geq 3$, $j=1,2,3$.  Thus jammed partitions are possible only for
networks of size $N=m_1+m_2+m_3\geq 9$.  Following the rules in
Eq.~(\ref{m123}), the following partitions are jammed: for $N=3k$ with $k\geq
3$, partitions of the form $(k,k,k)$; for $N=3k+2$ with $k\geq 3$,
$(k,k+1,k+1)$; and for $N=3k+1$ with $k\geq 4$, $(k,k,k+1)$.  Thus jammed
partitions indeed exist for $N=9$ and $N\ge11$.


Finally, we show that jammed states are impossible for $N\le8$ and $N=10$.
As a preliminary, we need the following:

\smallskip\noindent 
{\bf Lemma}.  Let $ABC$ be imbalanced, $ADC$ 
balanced, and $s_{AC}=-1$.  Then one of the two triads $ABD$ and $BDC$ is
balanced, and the other is imbalanced. 

\smallskip\noindent 
{\bf Proof}. Let $s_{XYZ}=s_{XY}s_{YZ}s_{ZX}$ be the sign of the triad $XYZ$.
For the imbalanced $ABC$ triad we have $s_{ABC}=-1$ while for the balanced 
triad $s_{ACD}=1$. Using additionally the identities $s_{XY}=s_{YX}$,
$s_{XY}^2=1$, we obtain the product of the signs of triads $ABD$ and $BDC$:
\begin{eqnarray*}
s_{ABD}s_{BDC}&=&s_{AB}s_{BD}s_{DA}s_{BD}s_{DC}s_{CB}\\ 
             &=&s_{AB}s_{DA}s_{DC}s_{CB}\\ 
             &=&s_{AB}s_{BC}s_{CA}s_{AC}s_{CD}s_{DA}\\
             &=&s_{ABC}s_{ACD}=-1
\end{eqnarray*}
from which the lemma immediately follows. 

Now suppose that there is a jammed state in a network with an even number of
nodes $N=2k$.  By definition, there is at least one imbalanced triad in the
jammed state; let $ABC$ be such an imbalanced triad with $s_{AC}=-1$.  Since
the state is stable, out of the $N-2=2k-2$ triads attached to the link $AC$,
at least $k$ are balanced.  (Note that this construction requires $k\ge 3$;
however, it is trivial to show that there is no jammed state for the case
$k=2$ [$N=4$]).  Take $k$ such balanced triads and denote them $AD_jC$,
$j=1,\dots,k$.  To each pair of triads $ABC$ and $AD_jC$ we now apply the
lemma.  Then there is a certain number $x$ of imbalanced triads among
$ABD_j$, and a certain number $y$ of imbalanced triads among $CBD_j$, with
$x+y=k$.  Stability ensures that there are at most $k-2$ imbalanced triads
attached to the link $AB$.  Recalling that $ABC$ is imbalanced and that there
are $x$ imbalanced triads $ABD_j$, we obtain $x+1\leq k-2$.  A similar
argument applied to link $CB$ leads to $y+1\leq k-2$.  Summing these
inequalities and using $x+y=k$ gives $k\geq 6$, or $N\geq 12$ for even
$N$.

The case of odd $N$ is similar.  We set $N=2k-1$, with $k\ge3$.  Now
each link is attached to at least $k$ balanced triads and at most $k-1$
imbalanced triads.  Repeating the same argument as for the even case we
obtain the conditions $x+1\le k-1$ and $y+1\le k-1$ with $x+y=k$, which leads
to $N\ge 9$ for odd $N$.

\subsection{Number of Jammed Configurations} 
\label{enumer}

Last, we show that the number of jammed configurations greatly exceeds the
number of balanced configurations.  The total number of distinct network
configurations is $2^L$.  Each balanced state has the form
$(m_1,m_2|\,m_1+m_2=N)$, and we enumerate all classes of balanced states by
counting the integer solutions of $m_1+m_2=N$, with $0\leq m_1\leq m_2$.
Therefore the number of classes of balanced states is $B=k+1$ for $N=2k$ and
$N=2k+1$.  The total number $\mathcal{B}$ of balanced states is determined
from
\begin{equation}
\label{balanced}
\mathcal{B}=\sum_{m_1+m_2=N}\binom{m_1+m_2}{m_1}=2^N ~,
\end{equation}
and is thus much larger than the number of classes of balanced states.

For the number of classes of jammed states $J$ and the number of jammed
states $\mathcal{J}$, we can only establish lower bounds.  For large $N$,
instead of exact counting we employ a continuum description.  From
Eq.~(\ref{m123}), the number of jammed partitions is equal to $N^2$ times the
area $\mathcal{A}$ of the region inside the triangle $x_1+x_2+x_3\leq 1$
defined by inequalities $x_1+x_2\geq x_3$, $x_2+x_3\geq x_1$, $x_3+x_1\geq
x_2$; this area is $\mathcal{A}=1/8$.  We also divide by $3!=6$ to account
for over-counting different permutations of $m_1, m_2, m_3$.  Thus
$J>N^2/48$.  We could improve this bound by counting additional jammed states
built from the construction in Fig.~\ref{strange9z}(c), but this contribution
would not affect the $N$ dependence of the bound.  However, we do not know
whether $J\propto N^2$ or $J$ grows faster than $N^2$ due to the existence of
jammed states, such as those in Fig.~\ref{strange9z}(a), that are not in the
classes described in Sec.~\ref{construct}.

We obtain a lower bound for $\mathcal{J}$ in a similar manner to that in
Eq.~(\ref{balanced}) for counting the number of balanced states,
\begin{equation}
\label{jammed}
\mathcal{J}>\sum_{\rm jammed}\binom{m_1+m_2+m_3}{m_1,m_2,m_3}\approx 3^N ~,
\end{equation}
where the sum is over jammed partitions, and the trinomial coefficient is
\begin{equation*}
\binom{m_1+m_2+m_3}{m_1,m_2,m_3}=\frac{(m_1+m_2+m_3)!}{m_1!\,m_2!\,m_3!} ~.
\end{equation*}
The summand in Eq.~(\ref{jammed}) is sharply peaked around $m_1=m_2=m_3=N/3$
and therefore the sum is very close to $3^N$ which is the sum over {\em all}
partitions.  Again, the lower bound may be weak because of the neglect of
non-tripartite jammed configurations.

In summary, we find that $\mathcal{J}>3^N\gg 2^N=\mathcal{B}$.  Thus the
total number of jammed states greatly exceeds the total number of balanced
states.  Nevertheless, for a random initial condition, the probability to end
in a balanced state is very close to one while the probability to end in a
jammed state is negligible; that is, the basin of attraction of balanced
states greatly exceeds the basin of attraction of jammed states.

\section{Summary and Discussion}

In social relations, we may encounter the uncomfortable situation of an
imbalanced triad.  If you have two friends that develop a mutual animosity,
then an imbalanced triad of relations exists.  You will then likely have to
choose between these two friends, thereby resolving the social conflict and
restoring the relationship triads to balance.  In this work, we implemented
simple and prototypical dynamical rules for healing imbalanced triads and we
investigated the resulting evolution of these social networks.  

In the case of local triad dynamics, a finite network falls into a
socially-balanced state, where no frustrated triads remain.  The time to
reach this final state depends very sensitively on the propensity $p$ for
forming friendly links in the update events that heal social imbalance.  For
an infinite network, the balanced state is never reached when $p<1/2$ and the
system remains in a stationary state.  The density of unfriendly links
gradually decreases and the network undergoes a dynamical phase transition to
an absorbing, paradise state for $p\geq 1/2$.

We also examined the dynamics in which an additional global constraint is
imposed that the number of imbalanced triads in the entire network cannot
increase in an update event.  The virtue of this dynamics is that the final
outcome is always reached quickly.  A downside, however, is that the final
configuration of the network may be jammed---these are states that are not
balanced, but where flipping any link increases the number of imbalanced
triads.  Fortunately, the probability of reaching a jammed state is
vanishingly small and the final state is either a two-clique bipolar state or
paradise.

As alluded to in the introduction, a natural application for social balance
ideas is to international relations, with the prelude to World War I being a
particularly appropriate example.  For example, the Three Emperors' League
(1872, and revived in 1881) aligned Germany, Austria-Hungary, Russia, leaving
France isolated.  However, the Turkish-Russian war (1877) and tension between
Austria-Hungary and the Balkan states unraveled Russia's participation in the
League, and a bipartite agreement between Germany and Russia lapsed in 1890.
In the meantime, the Triple Alliance was formed in 1882 that joined Germany,
Austria-Hungary, and Italy into a bloc that continued until World War I.

On the other hand, a French-Russian alliance was formed over the period
1891-94 that ended France's diplomatic isolation with respect to the Triple
Alliance.  Subsequently an Entente Cordiale between France and Great Britain
was consummated in 1904, and then a British-Russian agreement in 1907, after
long-standing tensions between these two countries, that then bound France,
Great Britain, and Russia into the Triple Entente.  While our historical
account of these Byzantine maneuvers is very incomplete (see Refs.~\cite{L}
for more information), the basic point is that among the six countries that
comprised the two major alliances, bipartite relationships changed as triads
became unbalanced and there was a reorganization into a balanced state of the
Triple Alliance and the Triple Entente that became the two main protagonists
at the start of World War I.

On the theoretical side, there are several avenues for additional research.
One possibility is to relax the definition of imbalanced somewhat.  This is
the direction followed by Davis \cite{Davis} who proposed the ``clusters
model'' in which triads with three unfriendly relations are deemed
acceptable.  The clusters model thus allows for the possibility that ``an
enemy of my enemy may still be my enemy.''  This more relaxed definition for
imbalanced triads may lead to interesting dynamical behavior that will be
worthwhile to explore.

Another natural generalization of the balance model would be to ternary
relationships of positive $+$, negative $-$, or indifferent $0$.  These
relations may lead to the emergence of cliques (groups of mutual friends who
dislike other people) and communities (groups of mutual friends with no
relations with other people).  It would be interesting to study the number of
cliques and number of communities as a function of network size and the
density of indifferent relationships.  Communities on the Web can be
effectively identified \cite{adamic}, and these results may allow for useful
comparisons between data and model predictions.  Finally, relations need not
be symmetric, that is, $s_{ij}$ may be different from $s_{ji}$, and it may be
interesting to generalize the basic notions of balance to networks with such
asymmetric interactions.

\acknowledgments{ TA gratefully acknowledges financial support from the Swiss
  National Science Foundation under the fellowship 8220-067591.  SR thanks K.
  Kulakowski for informative discussions about social balance, and also
  acknowledges financial support from DOE grant W-7405-ENG-36 (at LANL) and
  NSF grant DMR0227670 (at BU).}

\appendix

\section{Fluctuations in Local Triad Dynamics}
\label{fluctsig}

In this appendix, we compute the normalized variance $\sigma=\langle
\eta^2\rangle$.  We focus on the most interesting case of the critical regime
$p=1/2$, where fluctuations exhibit the asymptotic behavior of
Eq.~(\ref{sigma}).  Then we briefly discuss the two regimes $p<1/2$ and
$p>1/2$.

We first note that $A$ changes according to
\begin{equation}
\label{A-rule}
A\to
\begin{cases}
A-1   &{\rm rate}\quad N_3 \\       
A-1   &{\rm rate}\quad pN_1\\
A+1   &{\rm rate}\quad (1-p)N_1
\end{cases}
\end{equation}
which describe the processes $N_3\to N_2$, $N_1\to N_0$, and $N_1\to N_2$,
respectively.  {}From (\ref{A-rule}) we obtain
\begin{equation}
\label{Ap-eq}
\frac{d}{dt}\,\langle A\rangle=-\langle N_3\rangle-p\langle N_1\rangle
+(1-p)\langle N_1\rangle\,, 
\end{equation}
which simplifies to
\begin{equation}
\label{A-eq}
\frac{d}{dt}\,\langle A\rangle=-\langle N_3\rangle
\end{equation}
at the threshold $p=1/2$.  Since $\langle A\rangle\propto a$ and $\langle
N_3\rangle\propto \langle A^3\rangle\propto a^3$ to lowest order,
Eq.~(\ref{A-eq}) can be written as
\begin{equation}
\label{a-eq}
\frac{d a}{dt}=-a^3
\end{equation}
whose solution is given by Eq.~(\ref{rho-sol-crit}).  Similarly from
(\ref{A-rule}) we obtain
\begin{eqnarray}
\label{Ap2-eq}
\frac{d}{dt}\,\langle A^2\rangle&=&
\langle(-2A+1)N_3\rangle+p\langle (-2A+1)N_1\rangle\nonumber\\
&+&(1-p)\langle (2A+1)N_1\rangle
\end{eqnarray}
which, for $p=1/2$, simplifies to
\begin{equation}
\label{A2-eq}
\frac{d}{dt}\,\langle A^2\rangle=\langle N_3\rangle+\langle N_1\rangle
-2\langle AN_3\rangle\,.
\end{equation}
Subtracting Eq.~(\ref{A-eq}) multiplied by $2\langle A\rangle$ from
Eq.~(\ref{A2-eq}), we obtain the evolution equation for the variance
\begin{equation*}
\frac{d}{dt}\,\left(\langle A^2\rangle-\langle A\rangle^2\right)
=\langle N_3\rangle+\langle N_1\rangle+2\left(\langle A\rangle \langle N_3\rangle
-\langle AN_3\rangle\right).
\end{equation*}
Now using standard methods \cite{vK} to compute moments of the stochastic
variable $A=La+\sqrt{L}\eta$ (\ref{expansion}), we obtain the leading
behavior $\langle A^2\rangle-\langle A\rangle^2\propto \sigma$, while
$\langle N_3\rangle\propto a^3$, and $\langle N_1\rangle\propto 3a(1-a)^2$.
Similarly, the leading terms in $\langle A\rangle \langle N_3\rangle$ and
$\langle AN_3\rangle$ cancel, while the next correction is
\begin{equation}
\label{last}
\langle A\rangle \langle N_3\rangle-\langle AN_3\rangle \propto 3a^2\sigma-6a^2\sigma ~.
\end{equation}
Using these results for the various moments, the variance satisfies
\begin{equation}
\label{s-eq}
\frac{d \sigma}{dt}=-6a^2\sigma+a^3+3a(1-a)^2 ~.
\end{equation}
Dividing (\ref{s-eq}) by (\ref{a-eq}) we obtain
\begin{equation}
\label{sa-eq}
\frac{d \sigma}{da}=\frac{6}{a}\,\sigma-\frac{a^3+3a(1-a)^2}{a^3} ~.
\end{equation}
Since $\sigma=a^6$ solves the homogeneous equation $d\sigma/da=6\sigma/a$, we
seek a solution of the inhomogeneous equation (\ref{sa-eq}) in the form
$\sigma=a^6 s(a)$.  Equation (\ref{sa-eq}) becomes
\begin{equation*}
\frac{d s}{da}=-\frac{a^3+3a(1-a)^2}{a^9}=-\frac{3}{a^8}+\frac{6}{a^7}-\frac{4}{a^6} ~,
\end{equation*}
whose solution is 
\begin{equation}
\label{s-sol}
s=\frac{3}{7}\,\frac{1}{a^7}-\frac{1}{a^6}+\frac{4}{5}\,\frac{1}{a^5}+C ~.
\end{equation}
Thus $\sigma=a^6 s(a)$, with $s(a)$ given by (\ref{s-sol}).  The integration
constant $C$ is fixed to satisfy the initial condition $\sigma(a_0)=0$.  In
particular, for a totally antagonistic initial network ($\rho_0=0$ or
$a_0=1$), $C=-8/35$.  In this case $a=1/\sqrt{1+2t}$, and the variance
becomes
\begin{equation}
\label{sigma-sol}
\sigma=\frac{3}{7}\,\frac{1}{a}-1+\frac{4}{5}\,a-\frac{8}{35}\,a^6 ~.
\end{equation}

The leading asymptotic behavior $\sigma\to (3\sqrt{2}/7)\,t^{1/2}$ holds
independent of the initial condition.  Hence we establish the crucial result
(\ref{sigma}), which leads to the asymptotic behavior (\ref{correct}) for the
absorption time for $p=1/2$.

For $p\ne 1/2$, or $\epsilon=1-2p\ne 0$, we recast Eq.~(\ref{Ap-eq}) to
\begin{equation}
\label{ap-eq}
\frac{d a}{dt}=3\epsilon a(1-a)^2-a^3
\end{equation}
which is of course identical to Eq.~(\ref{rho}).  Following the same steps
that led to Eq.~(\ref{s-eq}), we then derive for the variance
\begin{equation}
\label{sp-eq}
\frac{d \sigma}{dt}=a^3+3a(1-a)^2-6\sigma[a^2-\epsilon(1-a)(1-3a)]\,.
\end{equation}

When $p<1/2$, both $a$ and $\sigma$ quickly approach stationary values
\begin{equation}
\label{as-inf}
a_\infty=\frac{\sqrt{3\epsilon}}{\sqrt{3\epsilon}+1}~,\quad      
\sigma_\infty=\frac{3}{4}\,
\frac{1}{\sqrt{3\epsilon}}\,\frac{1+\epsilon}{(\sqrt{3\epsilon}+1)^2}      
\end{equation}
The density of unfriendly links in a finite system therefore exhibits
fluctuations of the order of $\sqrt{\sigma_\infty/L}$ about the average
density $a_\infty$.  Close to the phase transition point, the magnitude of
fluctuations eventually becomes comparable with the average.  {}From
$a_\infty\sim \sqrt{\sigma_\infty/L}$ and Eq.~(\ref{as-inf}), we find that
this equality occurs when $\epsilon\sim N^{-4/3}$; this gives an estimate of
the width of the phase transition region due to finite-size effects.

When $p>1/2$, both $a$ and $\sigma$ vanish as $t\to\infty$.  A straightforward
asymptotic analysis of Eq.~(\ref{sp-eq}) yields 
\begin{equation}
\label{s-asymp+}
\sigma\to\frac{a}{|\epsilon|}\equiv \frac{a}{2p-1}\quad{\rm as}\quad a\to 0. 
\end{equation}
Fluctuations become comparable with the deterministic part when
$a\sim\sqrt{\sigma/L}\sim \sqrt{a/(2p-1)L}$, that is, when $La\sim
(2p-1)^{-1}$. Using $a\sim e^{-3(2p-1)t}$ from Eq.~(\ref{appr-to-paradise})
we estimate the time to reach paradise to be
\begin{equation}
\label{TN-above-impr}
T_N\sim (2p-1)^{-1}\,\ln[(2p-1) N^2].
\end{equation}
The difference between this result and Eq.~(\ref{TN-above}), which was
established using the naive criterion $La\sim 1$, is small because the factor
$2p-1$ appears inside the logarithm.


\end{document}